\documentclass[preprint,showpacs,preprintnumbers,amsmath,amssymb]{revtex4}

\usepackage{graphicx}
\usepackage{dcolumn}
\usepackage{bm}

\begin{document}

\preprint{}

\title{A link between ECM plasticity and synaptic morphological evolution}

\author{ S. Kanani$^1$, I.Robbins$^2$, T.Biben$^3$ and  N.Olivi-Tran$^{3,4}$} 
\affiliation{$^1$ Institut de Genomique Fonctionnelle, 141 rue de la Cardonille
34094 Montpellier cedex 05, France,\\ $^2$ Institut de Genetique Moleculaire de Montpellier, UMR-CNRS 5535
1919, route de mende
34293 Montpellier Cedex 5 , France \\$^3$Laboratoire de Physique de la Mati\`ere Condens\'ee et Nanostructures, UMR-CNRS 5586, Universit\'e Claude Bernard Lyon I, Domaine Scientifique de la Doua, 69622 Villeurbanne cedex, France \\ 
$^4$Laboratoire de Sciences des Proc\'ed\'es C\'eramiques et Traitements de Surface, UMR-CNRS 6638, Ecole Nationale 
Sup\'erieure de C\'eramiques Industrielles, 47 avenue Albert Thomas, 87065 Limoges cedex,
France }

\date{today}

\begin{abstract}
We made a link between Extra Cellular Matrix (ECM) plasticity and the morphological changes in synapses
after synaptic excitation. A recent study by Zhang et al \cite{zhang} showed that transmembrane voltage
causes movement of the cell membrane. Here we will study the relation between the mechanical properties
of collagen which is the major component of the ECM and synaptic morphological changes in relation with the theory of 
DO Hebb \cite{hebb}.

\end{abstract}

\vfill
\maketitle
\section{Introduction}
Hebb \cite{hebb} suggested that the representation of an object implies the activity of all cortical
cells activated by this stimulus. It is this group of neurons that Hebb called a cellular assembly.
Hebb thought that all cells were related by reciprocal connections. Hebb suggested also that
if the activity of this assembly of neurons had a sufficiently long duration, a consolidation of the 
information happened through a process which rendered these connections more efficient.
The memory effect was obtained (always following Hebb) by this 'preferential path'
(cellular assembly), i.e. if a part only of this preferential path was activated, the whole
cellular assembly would be stimulated, giving place to a memory effect.

This memory effect is linked to synaptic plasticity, on one hand.
Synaptic plasticity is linked to dynamic modulation of the postsynaptic membrane \cite{luscher}.
Moreover, long term enhancement of synaptic efficacy may lead to model learning and memory
\cite{bliss,engert}.
Luscher et al. \cite{luscher} observed  clear shape changes in denditric spines of dissociated
hippocampal cells \cite{fischer}. Indeed, they saw that spines seemed to oscillate
with a period of tens of seconds but did not seem to change their cross section. 

Straightforwardly, a recent study \cite{zhang} showed that for in vitro cells, transmembrane voltage modulates membrane tension
and therefore causes movement.
This study showed that the lateral displacement (perpendicular to the cell membrane) could be up to
$9 nm$ with a period of voltage pulses of $50 ms$. We can make a parallel between this movement
artificially created and oscillations of spines observed by Luscher et al \cite{luscher}.
The amplitude of the lateral movement  of the neurons in vivo depends on the electro membrane potential
of the neurons and therefore on ionic concentration in and out of the cell \cite{zhang}.

On another hand, Pizzo et al. \cite{pizzo} studied the relation between the mechanical properties of ECM
and the shape of cells, depending on the density in collagen fibrils and the duration of growth of the cells.
Another study \cite{kawano} showed that transected axons in the lateral hypothalamus
of mice could extend after the lesion if the deposition of type IV collagen and the formation
of a fibrotic scar was prevented. A similar article \cite{klapka} leaded to the same conclusion.

\section{Model}
The lateral movement due to transmembrane voltage modulations 
will create a strain on the collagen which surrounds the synapse
(here we deal indifferently with the axons and the dendrites) \cite{zhang}.

In brain, the neurons are embedded in an extra cellular matrix which is essentially composed
of type I collagen. This collagen matrix may be seen as isotropic: the collagen fibers have a random orientation.
But what will happen if the neurons apply a strain on this collagen matrix due to their lateral
displacement during transmembrane voltage modulations?

Therefore, a study by Feng et al. on the mechanical and rheological properties of collagen \cite{feng} deduced, from their
experimental results, the following equation for stress strain response of collagen:
\begin{equation}
\sigma = E_e. \gamma^2/(\gamma^2+b.\gamma+c)+ (\eta / \gamma) \sum_{\gamma_0}^{\gamma} e^{-[t(\gamma)-t(\gamma ')]/\lambda} d\gamma'
\end{equation}
where $\sigma$ is the stress, $\gamma$ the strain, $t$ the deformation time of  collagen, $\eta$ the sliding viscosity and $\lambda$
the relaxation time of collagen. The first term of right hand side of equation (1) is a non linear elastic plastic
term for collagen where $E_e,b,c$ are constants. The study of Feng et al \cite{feng} show that under strain,
collagen undergoes contraction and has a visco plastic behavior well described by equation (1) and in good agreement
with experiments.
 If the strain tends to a high value the stress will
tend straightforwardly to a constant. That is the plastic behavior of collagen.
Unrecoverable strain in cyclic loading test on collagen resulted in cyclic creep \cite{feng}.
We can make a comparison with cyclic loading and cyclic normal strain due to lateral cyclic movements of neurons 
during the propagation of an electro membrane potential.

Let us analyse the behavior of collagen next to the membrane or next to a synapse.
For that we have to evaluate the values of $\sigma$, and therefore $\gamma$,$t$,$\eta$,$\lambda$ and $Ee,b,c$.
From literature $t$ is equal to $1ms$ to $50ms$ \cite{neurosciences}, and $\lambda$ is equal to $30min$
to $1h$ \cite{feng}. In order to obtain $\eta$ and $\gamma$ one has to make a dimensional analyzis of these
two parameters.
$\gamma$ is a force applied by the membrane on the collagen, its dimensions are $kg.m.s^{-2}$.
Therefore, in terms of characteristics of the collagen which mass is $24.10^{-23}kg$ for a cube of 1000 molecules of lysine
(for type I collagen), which length is of the order of $10nm$ for 10 molecules of lysine and taking the characteristic
time equal to  the deformation time, we obtain $\gamma=24.10^{-25}N$.
With the same reasoning, $\eta$ is a viscosity thus its dimensions are $kg.m^{-1}.s^{-1}$ and its value is
equal to $24.10^{-12}kg.m^{-1}.s^{-1}$ .

In order to study the results obtained for values of $E_e,b,c=1$, we obtained the stress $\sigma$
following equation (1) approximatively equal to $10^{13}kg.m^{-1}.s^{-1}$
which i as sufficient value to have an effect on the geometry of the collagen
or ECM near the dendrite or axone.

\section{Discussion}
Now that preferential paths have been created in our simple model of brain, let us also model
the memory effect. Ganguly-Fitzgerald et al \cite{ganguly} showed that for Drosophilia,
exposure to enriched environments (i.e. exposure to rich sensorial environments)
affects the number of synapses and the size of regions involved in information processing \cite{praag,heisenberg}.
In our model, the preferential paths in the extra cellular matrix are regions where
there is a lower concentration of collagen close to  the neuron. Therefore, it will be easier for
two neurons to connect along these preferential paths by a simple steric effect.
Indeed, if we suppose that the direction of growth of neurons (dendrites or axons) is simply
leaded by the stiffness of the extracellular matrix, the connection of two neurons via
one synapse will be more frequent on the preferential paths.
Plus, during paradoxical sleep,
neurons undergo random lateral vibrations due to random propagations of neuronal excitations.
The memory of an event (e.g. sensorial) will be the creation of a preferential path plus
the creation of new synapses on this preferential path. 

Memory and intelligence are linked: intelligence is the ability to link two different
events which have been put into memory. Once again, if paradoxical sleep corresponds
to a random propagation of neuronal excitation, the possibility to link two different
preferential paths is linked the locations of these preferential paths and on the intensity
of neuronal excitation during sleep.

\section{conclusion}
To conclude,  for young mammals, the water concentration of brain is larger and therefore
the sliding viscosity $\eta$ and the relaxation time $t$ of collagen based extra cellular matrix
will be smaller than for the corresponding adult mammals. Therefore the stress $\sigma$ resulting from
the lateral vibrations of neurons will be smaller (see equation (1)).
This will lead to vanishing preferential paths and this explains the lack of long term memory
in young mammals.


\begin{thebibliography}{99}
\bibitem{hebb}{D.O. Hebb {\it The Organization of Behavior: A Neuropsychological Theory} (Wiley, New York, 1949)}
\bibitem{luscher}{ C.L\"uscher, R.A.Nicoll, R.C.Malenka and D.Muller, Nature Neuroscience, {\bf 3} (2000) 545}
\bibitem{bliss}{T.V.P.Bliss and G.L. Collingridge, Nature {\bf 361} (1993) 31}
\bibitem{engert}{F.Engert and T.Bonhoeffer, Nature {\bf 399} (1999) 66}
\bibitem{fischer}{M.Fischer, S.Kaech, D.Knutti and A.Matus, Neuron {\bf 20} (1998) 847}
\bibitem{zhang}{P.C. Zhang, A.M.Keleshian and F.Sachs, Nature {\bf 413} (2001) 428}
\bibitem{pizzo}{A.M.Pizzo, K.Kokini, L.C.Vaughn, B.Z. Waisner and S.L. Voytik-Harbin, J. of Appl. Physiol. {\bf 98}
(2005) 1909}
\bibitem{kawano}{H.Kawano, H.P.Li, K.Sango, K.Kawamura and G.Raisman, J. of Neurosci. Res. {\bf 80} (2005) 191}
\bibitem{klapka}{N.Klapka and H.W. M\"uller, J. of Neurotrauma {\bf 23} (2006) 422}
\bibitem{neurosciences}{M.F.Bear, B.W. Connors, M.A. Paradiso, {\it Neuroscience: Exploring the Brain}
Second Edition, (Lippincott Williams \& Wilkins Eds. , Baltimore , USA, 2001)}
\bibitem{feng}{Z.Feng, M.Yamato, T.Akutsu, T.Nakamura, T.Okano and M.Umezu, Artif. Organs {\bf 27} (2003) 84}
\bibitem{ganguly}{I. Ganguly-Fitzgerald, J.Donlea and P.J.Shaw, Science {\bf 313} (2006) 177}
\bibitem{praag}{H. van Praag, G.Kempermann and F.H.Gage, Nat. Rev. Neurosci. {\bf 1} (2000) 191}
\bibitem{heisenberg}{M.Heisenberg, M.Heusipp and C.Wanke, J. Neurosci. {\bf 15} (1995) 1951}

\end{thebibliography}
\end{document}